# On-chip optical diode based on silicon photonic crystal heterojunctions

Chen Wang, Chang-Zhu Zhou and Zhi-Yuan Li∗

Laboratory of Optical Physics, Institute of Physics, Chinese Academy of Sciences, P. O. Box 603, Beijing 100190, China

Optical isolation is a long pursued object with fundamental difficulty in integrated photonics. As a step towards this goal, we demonstrate the design, fabrication, and characterization of on-chip wavelength-scale optical diodes that are made from the heterojunction between two different silicon two-dimensional square-lattice photonic crystal slabs with directional bandgap mismatch and different mode transitions. The measured transmission spectra show considerable unidirectional transmission behavior, in good agreement with numerical simulations. The experimental realization of on-chip optical diodes using all-dielectric, passive, and linear silicon photonic crystal structures may help to construct on-chip optical logical devices without nonlinearity or magnetism, and would open up a road towards photonic computers.

PACS numbers: 42.79.-e, 42.25.Gy, 42.82.Gw

Optical isolation is a long pursued object with fundamental difficulty in integrated photonics [1]. The need to overcome this difficulty is becoming increasingly urgent with the emergence of silicon nano-photonics [2-4], which promises to create on-chip large-scale integrated optical systems. Motivated by the one-way effect, considerable effort has been dedicated to the study of the unidirectional nonreciprocal transmission of electromagnetic waves, showing important promise in optical communications. These "optical diodes" include fluorescent dyes with a concentration gradient [5], absorbing multilayer systems [6], and second harmonic generators with a spatially varying wave vector mismatch [7]. An electro-tunable optical isolator based on liquid-crystal heterojunctions, showing nonreciprocal transmission of circularly polarized light in photonic bandgap regions, has been reported [8]. In another configuration using liquid crystals, linearly polarized light is used [9]. In addition to many attempts on magneto-optical materials [10-13], optical isolators have also been fabricated using nonlinear optical processes [14,15] and electro-absorption modulators [16]. A theoretical scheme of optical isolation achieved by dynamically inducing indirect photonic transitions in an appropriately designed photonic structure was reported [17]. Other reported optical isolators are generally based on photonic crystals (PCs), which are dielectric or metallo-dielectric structures with a spatial periodicity in their refractive index [18,19]. This periodicity affects the propagation of electromagnetic waves in the same way as the periodic potential in a semiconductor crystal affects electron motion by defining allowed photonic bands and forbidden bandgaps [19]. For example, a one-dimensional photonic crystal structure having a nonlinear optical response and a two-dimensional (2D) photonic crystal waveguide with an asymmetric array of nonlinear defect rods [20-22] are found to display nonreciprocal effects. In these cases, however, optical isolation occurs only for specific high power ranges or with associated modulation side bands. In addition, there have been works that aim to achieve partial optical isolation in reciprocal structures that have no inversion symmetry (for example, chiral structures). Until now, on-chip integration of optical diode still stays in theory, particularly in silicon.

An efficient routine to create optical diode is via time-reversal symmetry breaking

[5-17,20-22] or spatial inversion symmetry breaking [23]. Both these effects could lead to the optical isolation in any device where the forward and backward transmissivity of light is very much different. Recently a unidirectional sound propagation has been reported in sonic-crystal-based acoustic diodes with a completely linear system [24]. In this paper, we report a method for making unidirectional on-chip optical diodes based on the directional bandgap difference of two 2D square-lattice photonic crystals comprising a heterojunction structure and the break of the spatial inversion symmetry. Simulations confirm the existence of a clear isolation effect in the designed heterojunction structure. We fabricate these on-chip optical diodes in silicon and the near-infrared experiment results show high-performance optical isolation, in good agreement with the theoretical prediction. This device may play the same basic role as the electrical diode in photonic circuit and further lead us to the achievement of on-chip optical logical devices without nonlinearity or magnetism and bloom the photonic network integration.

Figure 1(a) shows the schematic configuration of the original diode structure under study, which consists of two PC slab domains ($PC_1$ and $PC_2$) with the same lattice constant $a$ but different air hole radii ($r_1$ and $r_2$, respectively) comprising a heterojunction structure. These two PC regions stand at a silicon slab (grey area in Fig. 1(a)). Each PC region has a square-lattice pattern of air holes (white holes in Fig. 1(a)), with the hetero-interface between $PC_1$ and $PC_2$ along the Γ-M direction. Here we set the two hole radii as a fixed ratio to the lattice constant $a$, which are $r_1=0.24a$ and $r_2=0.36a$ in order to simplify our discussion. These two composite PCs would comprise a pure PC region if $r_1=r_2$. The light source is placed symmetrically aside the structure with two $4a$-wide ridge waveguides connecting the surface of the two PC regions. The whole area is surrounded by a perfectly matched layer.

We simulated the transmission spectra for a TE-like light signal transporting along the forward (from left to right) and backward (from right to left) direction using the three-dimensional finite-difference time-domain (3D-FDTD) method [25]. The refractive index of the dielectric slab was set to 3.4, corresponding to that of silicon at 1,550 nm. The slab thickness was $h=0.5a$. Note that all geometric parameters are constant and normalized by the PC lattice constant a. Figure 1(b) shows the calculated forward (black line) and backward (red line) transmission spectra. The frequency is normalized by $a/\lambda$. It is clearly seen that there exists an isolation band ranging from 0.2649 to 0.2958 ($a/\lambda$), where the forward transmission forms a peak with a transmissivity of about 6% while the backward transmissivity is down between 0.5% and 1%. The forward peak is located at 0.2793 ($a/\lambda$), just in the middle of the isolation band. We define S=($T_F$－$T_B$)/($T_F$＋$T_B$) as the signal contrast of the diode, where $T_F$ and $T_B$ denote the forward and backward transmissivity, respectively. The maximum S of this original diode equals 0.846 at the peak. Besides, there exists another isolation region from 0.2196 to 0.2649 ($a/\lambda$), where the backward transmissivity is higher than the forward transmissivity. This structure thus shows an extraordinary phenomenon of unidirectional transport property. When we set the interface of the hetero-junction as normal to the Γ-X direction of both $PC_1$ and $PC_2$, the simulation result shows no isolation phenomenon. This indicates that the isolation happens only if the hetero-junction is tilted.

The TE-like electromagnetic field distributions of the forward and backward transport of light at 0.2793 ($a/\lambda$) are calculated in order to have a visual sight of the detailed characteristic of optical isolation (Figs. 1(c) and 1(d)). In Fig. 1(c), light goes straight in $PC_1$ and reaches the heterojunction interface. Then one part of light reflects at the junction and goes downward; the other part of light travels along the junction and diffuses into $PC_2$, which eventually outputs from $PC_2$ into the connected waveguide. But in Fig. 1(d), light cannot go straight but separates into two 45°-direction paths, which fit the Γ-M direction of a square-lattice photonic crystal. Most of light is lost and there is little light reaching the output waveguide. It is thus the different choice of light path that leads to the difference between the forward and backward transmission.

To have a deeper understanding of the isolation effect, we calculated the band diagram of the TE-like modes of these two PC slabs using the 3D-FDTD method. Figures 2(a) and 2(b) show the calculation results. The first band (even mode) in bulk PC2 (Fig. 2(b)) is directional as the top mode frequency in the Γ-X direction (x-axis) is 0.2345 ($a/\lambda$) but that in the Γ-M direction (45°-direction) is 0.3087 ($a/\lambda$). Inside the region between 0.2345 ($a/\lambda$) and 0.3087 ($a/\lambda$), the all-directional transparent region of PC1 needs to be above 0.2633 ($a/\lambda$) (Fig. 2(a)) in order to match the bottom mode frequency in the Γ-X direction of the second band (odd mode). These two modes in PCs are the basic working mode of the diode structure. Here the even and odd modes are defined with respect to the off-slab mirror-reflection symmetry $\sigma_z$ of the field component $E_y$. In the region between 0.2633 ($a/\lambda$) and 0.3087 ($a/\lambda$) $PC_1$ is transparent in all directions, while $PC_2$ is transparent along the Γ-M direction but opaque along the Γ-X direction. Compared with Fig. 1(b), the isolation region [0.2649 to 0.2958 ($a/\lambda$)] just coincides with the overlapped region between the directional bandgap of PC2 and the all-directional pass band of $PC_1$. This simple picture indicates that the current isolation effect involves two ingredients: (I) directional bandgap of $PC_2$ and (II) all-directional pass band of $PC_1$. Noting that the structure doesn't obey the spatial inversion symmetry alone the propagating direction, the principle of optical isolation can thus be summarized as follows:

1) Forward. When light goes across $PC_1$ as the odd mode and reaches the hetero-junction along the Γ-X direction, it cannot stay in the Γ-X direction in $PC_2$ further because of the Γ-X directional gap. But the hetero-junction is along the Γ-M direction, so light turns to the hetero-junction and diffracts as the even mode at any Γ-M direction into $PC_2$, which eventually outputs.

2) Backward. When light goes directly into $PC_2$ as the even mode, it turns to the two Γ-M direction paths which can't convert to the odd mode of $PC_1$ in the Γ-X direction and eventually leak out so that it doesn't output.

Another point that should be explained is that the main diode structure is a quasi-diffraction structure with two PCs heterojunction. The waveguides attached to the PCs is only for experiment convenience. The band diagram of the TE-like modes of these two PC slabs (Fig. 2) not only gives us the directional bandgap difference, but also shows that every frequency has only one mode distribution corresponding in the two PCs individually in the diode working frequency region. All other input/output modes from the waveguide should first coupling into/out of these two working modes, indicating the mode pattern doesn't affect the diode property. We further find that the output mode conversion is output-waveguide-width related and the only reason is that it's caused by the coupling between $PC_2$ and output waveguide. This indicates that we can change the type or the shape of the output waveguide to improve the coupling but the diode effect remains the same. And due to its modular design, the conversed-mode problem also can be solved by adding other modules. This character also shows the advantages in optical integration that it has no self-needed situation which could affect other devices and easy to integrate to other devices.

As we have known that it is the directional bandgap difference that causes the isolation effect, next we change the original diode into the revised structure (Fig. 3(a)) in order to ensure the backward transmission to be zero and increase the magnitude of the forward transmission peak as much as possible. In Fig. 3(b) the isolation region still ranges from 0.2625 to 0.3050 ($a/\lambda$). The simulation values of the backward transmission in the isolation region generally decrease a half compared with that of the original structure and at the same time the forward peak reaches about 13% of the input power at the frequency of 0.2834 ($a/\lambda$). The maximum signal contrast S of the revised structure increases to 0.92 at the peak, which is near the value of the present electronic diodes. In addition, the region where the backward transmissivity is higher than the forward transmissivity is gone as the structure is no longer y-symmetric. The TE-like electromagnetic field distributions of the forward and the backward transport of light signals are calculated (Figs. 3(c) and 3(d)). It is clear that light signals are all guided out of the structure in the backward path (Fig. 3(d)) due to the Γ-M directional bandgap of $PC_2$ in contrast to the strong coupling into the output waveguide in the forward path (Fig. 3(c)).

Based on the numerical analysis of the optical diode, the revised diode structures together with the original structures were fabricated in silicon. The patterns were first defined in resist using the electron beam lithography (EBL) on the top layer of a silicon-on-insulator (SOI) chip. The resist patterns were then transferred to silicon layer using the inductive coupled plasma reactive ion etching (ICP) technique. The lattice constant $a$ was set to 440 nm so that the radii $r_1$ and $r_2$ were approximately equal to 110 nm and 160 nm. The slab thickness was 220 nm. The insulator layer ($SiO_2$) underneath the silicon pattern regions was finally removed by a HF solution to form an air-bridged structure. Figures 4(a) and 4(b) show the scanning electron microscopy (SEM) images of the fabricated diode structures along the light path. Light from a semiconductor laser, which is tunable between 1,500 nm and 1,640 nm, was directly coupled into the photonic crystal heterojunction diode with the aid of tapered ridge waveguides in the input and output ends [26]. The wavelength region is normalized to 0.2683-0.2933 ($a/\lambda$), enough to encompass the isolation region.

Figures 4(c) and 4(d) show the theoretical and experimental results of the transmission spectra of the diodes in the forward and backward directions. In Fig. 4(c) the theoretical forward peak of the original diode structure is at 1,575 nm [0.2793 ($a/\lambda$)] and the maximum transmissivity is 6%. The experimental forward peak is at 1,556 nm and the maximum transmissivity is 7%. In Fig. 4(d) the experimental forward peak is at 1,534 nm and the maximum transmissivity is 10% for the revised diode, whereas the theoretical forward peak is at 1,552 nm [0.2834 ($a/\lambda$)] with a transmissivity of 13%. The measured signal contrast S equals 0.718 (the original structure) and 0.831 (the revised structure) at the peak frequency. Both experimental peaks in Figs. 4(c) and 4(d) have a near 20 nm shift and 50 nm broadening against the theoretical simulations, which is probably due to the imperfections in fabrication. The experiment confirms the existence of the isolation effect in agreement with the theoretical prediction. Due to the arbitrariness of the lattice constant a, we can freely adjust the isolation frequency to anywhere as desired. This could be more convenient for the design of realistic photonic devices.

The principle for optical diode as analyzed in the above is robust as it is based on a simple directional bandgap mismatch effect of photonic crystal heterojunction. Yet, it should be noticed that in Figs. 4(c) and 4(d) the backward transmissions are fluctuating within 1% to 2% and both are higher than the simulation values, as a result, the signal contrast S degrades from 0.846 (0.92) in theory to 0.718 (0.831) in experiment. The performance improvement of the diode relies on how to maximize the peak of the forward transmissivity and minimize the backward transmissivity in experiment. Several means can help improve the forward transmissivity. First, one can change the air hole size of $PC_1$ and $PC_2$ and enlarge the directional bandgap. Calculations show that the forward peak transmissivity of the revised structure with $r_1$ =0.30$a$ and $r_2$ =0.45$a$ grows up dramatically to 29.4% while maintaining the same low level of backward transmissivity. Second, one can change the relative size of the input and output waveguides. Calculations show that by changing the input waveguide width to 2$a$ and keeping the output waveguide width 4$a$, the forward peak signal increases up to 20.8% in transmissivity.

To reduce the backward transmissivity, one can either enlarge the directional bandgap of $PC_2$ to attenuate the backward signal more strongly, or eliminate the return of leak-out light from the outside of slab or the structure boundary by introducing the absorbing metal dots near the structure, or enlarge the heterojunction structure appropriately so that the leak-out light cannot enter the output waveguide. Following the above general ideas, we further optimize the optical diode structures as illustrated in Fig. 5. The structure has parameters of $r_1$=100 nm and $r_2$=160 nm, and the input and output waveguide width are 2$a$ and 6$a$, respectively. In experiment, we have got an optical diode with an average of 21.3% of the forward peak transmissivity and 0.885 of the signal contrast S at 1,557 nm. Yet, the backward signal intensity is still not reduced to the low level predicted by theory. The performance improvement eventually depends on the nanofabrication precision, as structural disorders or imperfections will induce light scattering to background and reduce the signal contrast.

Our photonic crystal heterojunction diode has advantages of high signal contrast, wavelength-scale small sizes, and being all-dielectric, linear, and passive. Furthermore, it has a much smaller scale than those based on diffraction gratings [23,24] and thus greatly facilitates large-scale integration. The high performance on-chip optical diode realized in silicon without nonlinearity or magnetism will stimulate the exploration of other more complex on-chip optical logical devices with ultra-high stability, integration and much less power consumption. Such an optical diode may play the same basic role as the electrical diodes, which have significantly revolutionized fundamental science and advanced technology in various aspects of our routine life due to their capability of rectification of current flux, in photonic circuit and its large-scale fabrication could be readily achieved by the well-developed CMOS techniques. The realization of high-performance on-chip optical diodes may open up a road toward photonic computers.

This work was supported by the National Natural Science Foundation of China at No. 10525419 and the State Key Development Program for Basic Research of China at No. 2007CB613205.

∗To whom correspondence should be addressed.

lizy@aphy.iphy.ac.cn

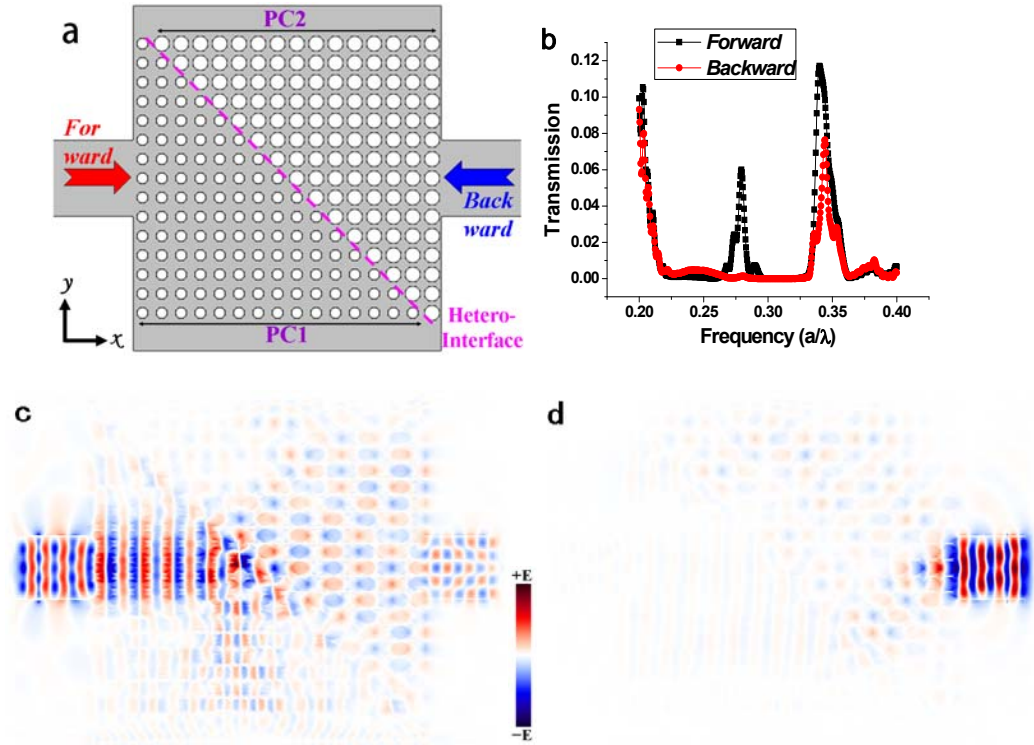

FIG. 1 (color online). (a) Schematic geometry of an original heterojunction optical diode formed by the interface (normal to the Γ-M direction) between two PC slabs (denoted as $PC_1$ and $PC_2$) with different hole radii ($r_1$ and $r_2$, respectively). (b) Simulated transmission spectra of the diode in the forward direction (the black line) and the backward direction (the red line). An input and output ridge waveguide has been used in the 3D-FDTD calculation. (c),(d) Calculated TE-like mode $E_y$ field distribution at 0.2793 ($a/\lambda$) in the forward and backward direction, respectively.

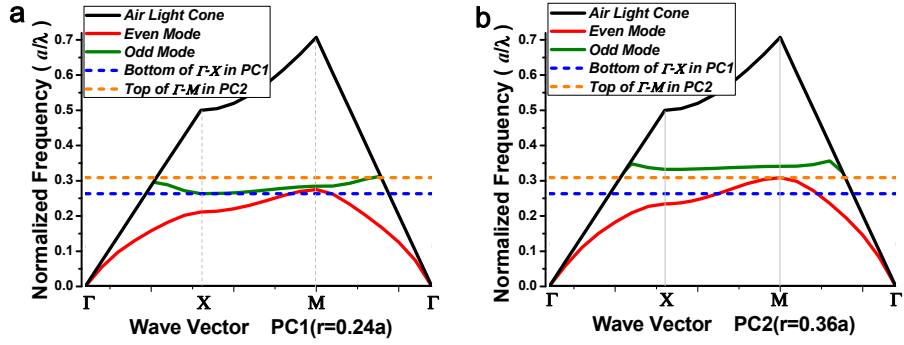

FIG. 2 (color online). (a) Calculated modal dispersion curve for $PC_1$ ($r=0.24a$). (b) Calculated modal dispersion curve for $PC_2$ ($r=0.36a$), in which the black line is the air light cone. The red curve denotes the first even mode, while the green curve denotes the second odd mode. The blue dashed line denotes the bottom frequency of the Γ-X directional odd mode [0.2633 ($a/\lambda$)] of $PC_1$ and the orange dashed line denotes the top frequency of the Γ-M directional even mode of $PC_2$ [0.3087 ($a/\lambda$)].

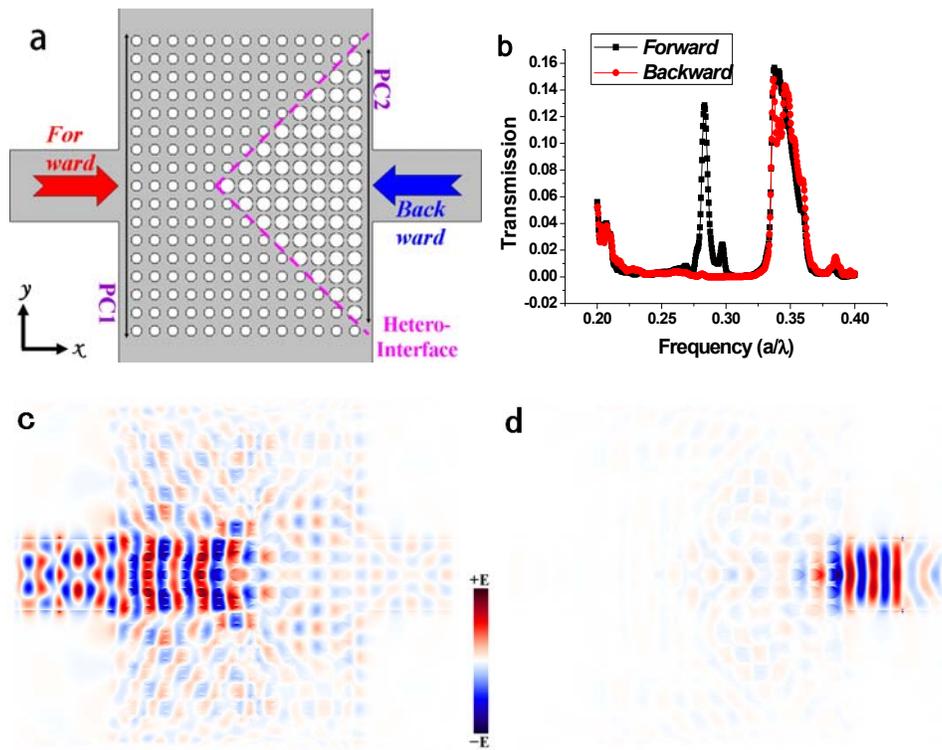

FIG. 3 (color online). (a) Schematic geometry of the revised diode structure formed by two heterojunction interfaces (normal to two Γ-M directions) between two PC slabs. (b) Simulated transmission spectra of the revised diode structure in the forward direction (the black line) and the backward direction (the red line). (c),(d) Calculated TE-like mode Ey field distribution at 0.2834 ($a/\lambda$) in the forward and backward direction, respectively.

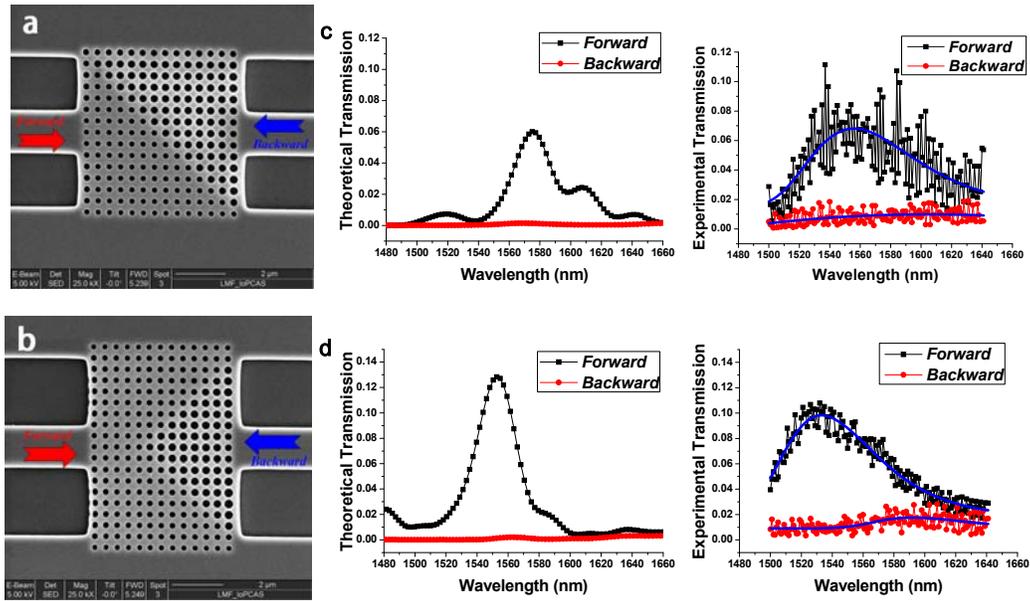

FIG. 4 (color online). (a),(b)Scanning electron microscope images of the original and the revised optical diode structures. (c),(d) Theoretical (left, part of Fig. 1(b)&3(b)) and experimental (right) transmission spectra of the original and the revised diode structure.

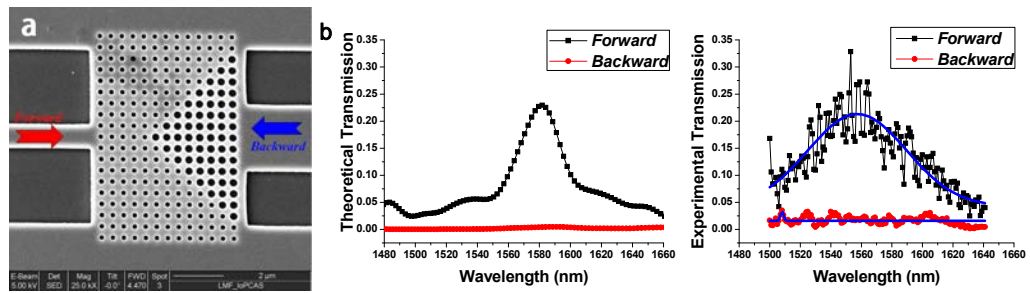

FIG. 5 (color online). (a) Scanning electron microscope images of the optimized optical diode system. (b) Theoretical (left) and experimental (right) transmission spectra of the optimized diode structure.